\documentstyle[aps,prl,twocolumn,epsfig]{revtex}

\begin{document}

   \draft


   \twocolumn[\hsize\textwidth\columnwidth\hsize\csname @twocolumnfalse\endcsname%


   \title{Pareto's law:  a model of human sharing and creativity}
   \author{Nicola Scafetta,$^{1}$  Sergio Picozzi$^{2}$ and Bruce J. West.$^{1,2,3}$}
   \address{$^{1}$ Pratt School of EE Department, Duke University,  P.O. Box 90291, Durham, NC 27701. }
   \address{$^{2}$ Physics Department, Duke University, Durham, NC 27701. }
   \address{$^{3}$ Mathematics Division, Army Research Office, Research Triangle Park, NC 27709-2211. }
    \date{\today}
   \maketitle

\begin{abstract}
A computational model for the distribution of wealth among the members of an
ideal society is presented. It is determined that a realistic distribution
of wealth depends upon two mechanisms: an asymmetric flux of wealth in
trading transactions that advantages the poorer of the two traders and a
non-stationary creation and destruction of individual wealth. The former
mechanism redistributes wealth by reducing the gap between the rich and
poor, leading to the emergence of a middle class. The latter mechanism,
together with the former one, generates a distribution of wealth having a
power-law tail that is compatible with Pareto's law.
\end{abstract}

   \vspace{0.5cm}
   %
   ]
   %


More than a century ago the Italian sociologist and economist, Vilfredo Pareto,
studied the distribution of income among people of different western
countries and found an inverse power law\cite{pareto}. The cumulative probability $P(w)$ of
people whose income is at least $w$ seemed to vary as $P(w)\propto
w^{-\alpha }$. Pareto mistakenly believed that the exponent $\alpha$ was  universal constant with an approximate value of $\alpha =1.5$\cite{montroll}. 
In our discussion we use the terms distribution of income and distribution
of wealth interchangeably, and although the two distributions may not
exactly coincide, they are in fact strongly dependent on one another. The
distinction between the two will not influence the conclusions drawn from
the dynamical models of economic interactions we present here.

Societies in the West have historically been partitioned into three classes;
the poor, the middle and the rich. The relative size of each class is
determined by somewhat arbitrarily assigned levels of income, but it is safe
to say that the smallest class is the rich and for a stable society the
largest is the middle class. This partitioning must be incorporated into any
mathematical model describing how wealth is distributed within a society.
The inverse power law of Pareto does not have this characteristic
partitioning, because the derivative of a power law is still a power law.
According to data, across the full range of income, we should expect the probability density function (pdf) $p(w)$  to increase at low income, reach a maximum and, finally, decrease with
increasing wealth. Moreover, fitting the cumulative probability $P(w)$
instead of the pdf $p(w)$ will certainly mask important properties of  the
dynamics of the economy. Therefore, herein we model the density function $p(w)$.
The cumulative probability $P(w)$ is related to the pdf $p(w)$ via
the relation $P(w)=\int _{w}^{\infty }p(x)dx$.

We investigate a computational model of the distribution of wealth, which is
based on the assumption that the pdf of wealth is the result of a
non-stationary total wealth and asymmetric trading mechanisms in the
economy. For a non-stationary economy we postulate a situation in which the
total wealth of a society is not conserved, since an individual may create
or consume wealth. For an asymmetric trading economy we postulate a
mechanism by which, through trades,  wealth randomly moves from one
individual to another, but in such a way as to, on average, slightly
advantage the poorer trader in the exchange.

Wealth decays because of human consumption and needs to be continuously
recreated through work and human creativity. People must work to live! The
mechanisms that regulate the creation and consumption of wealth are neither
identical nor symmetric, and they do not compensate exactly for one another.
This lack of compensation implies that the wealth of an individual, even in
the absence of any type of trade, fluctuates in time. Individual wealth may
increase or decrease according to the situation. Moreover, also the 
assumption that all transactions occur at precisely the right price, a fact
that implies the absence of any flux of wealth between traders, is
unrealistic. Some sellers are more persuasive and some buyers are more
discerning than others. There are good and bad deals. Therefore, the proper
value of an asset may be lower or higher than the price paid for it. This
dynamic of a transaction continuously generates a random flux of wealth from
one trader to the other in every transaction.

Finally, we assume that this random flux of wealth during a transaction
depends upon the difference in wealth between the two traders and may be
statistically biased toward the poorer partner. This mechanism seems to be
needed to redistribute the wealth among people.  More importantly, without
this mechanism, it would be true that the rich would get richer and the poor
would get poorer. But this latter flux of wealth from the poor to the rich
would be a consequence of the excess wealth of the rich and not the result
of any hypothetical economic abuses of the poor by the rich. In a fair game
in which the rich and the poor have the same probability to win or lose in
any transaction involving one of each class, the rich ultimately win because of
their larger resources \cite{feller}. So for a society to avoid the rich getting too rich
and the poor getting too poor with the fallout of a social collapse, the poor must be slightly advantaged in their transactions with the rich.

The {\it symmetry-breaking} mechanism that advantages the poor, in their
transactions with the rich,  should be thought
to be both a necessity for, and a consequence of, an efficient and stable
society. In fact, the necessity to slightly advantage the poor requires a
policy of rights and freedom, balanced by duties and responsibilities, that has been
typical of western societies. For example, in a modern democratic society the
right of the employed worker to be protected by a union is recognized, as is
their right to strike against their employers in particular cases. These
rights allow the workers to obtain a salary statistically higher than the
real value of their job, compatible to the actual economy. In the same way,
the freedom to decide the price of their own goods is not only an incentive
to a larger and better production of goods, but allows the self-employed of
the working class to statistically increase the price of their own products.
In fact, many such workers are farmers, fishermen and small scale artisans,
who produce the basic needs of a society, for example, food. Even the
richest person needs to eat, therefore, if he/she cannot enslave the farmer
and wants to eat, he/she must buy food at the price the farmer decides. We
understand that the economic situation is much more complex than we are able
to detail here, but the policy of rights and freedom in an economic
context contributes the symmetry-breaking mechanism in our model as we describe.

In the physics tradition we hypothesize a mathematical model describing the
economic mechanisms in an idealized situation, predict the resulting
distribution of wealth and then compare the predicted distribution function
with data \cite{stanley,badger}. The correctness of our assumptions are ultimately determined by
data and not logical discourse. In all our simulations we suppose an
economic society of 10,000 agents, who at the onset have the same degree of
wealth. We calculate the wealth pdf, $p(w)$, of our hypothetical society
after 100 million random transactions between all agents. We proceed by
steps, introducing models of increasing complexity and discussing the
consequences of each of them in turn. The first three models assume the
conservation of total wealth, while the last model assumes that the
total wealth of a society may change.

\textbf{Perfect trade price.} The simplest hypothesis is to suppose that
wealth cannot be created or destroyed and that all transactions between
agents occur at the right price. This complete symmetry means that each
agent acquires the exact same amount of wealth that is given up to the other
trader. The model is based upon the assumption that the price of an asset is
defined by its intrinsic value. Consequently the wealth of each agent
remains constant and the final distribution of a closed society's wealth
remains identical to the initial uniform distribution. This model is clearly
at odds with our experience of the economic world.
\begin{figure}
\epsfig{file=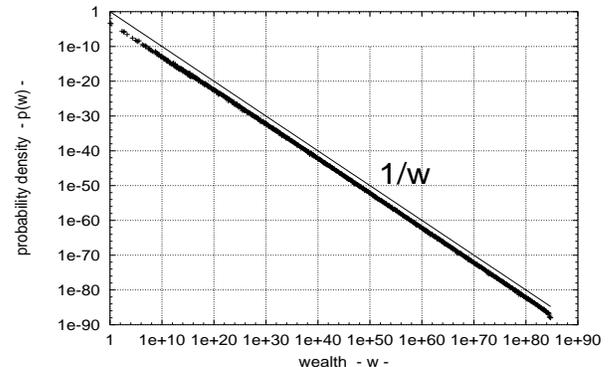,height=8cm,width=5cm,angle=-90}
\caption{Symmetric, stationary economy. The wealth is measured in  units of the poorest agent's wealth.  }
\end{figure}
\textbf{Random symmetric trade price.} Again we suppose that wealth cannot
be created or destroyed, however we now assume that the transactions between
agents may occur at a price that randomly fluctuates around a hypothetical 
{\it ideal} value. Therefore, the wealth of each agent may increase or decrease
according to whether he/she makes good or bad deals. We still assume
symmetry, that is, both agents have the same chance to win or lose, such as
in a coin toss. We also suppose that the maximum wealth that may move from
one agent to the other, in a single transaction, is limited to some fraction
of the wealth of the poorer of the two traders. This restrictive condition is
required because only in a robbery is it realistic to assume that a person
will give over to another more wealth than he/she possesses.

Fig. 1 shows the pdf of wealth among the agents after 100 million random
transactions. It is a very wide inverse power law of the type 
$p(w)\propto 1/w$. The  fraction of wealth that moves from one agent to the
other in a single transaction is assumed to vary between 0\% and 50\% of the poorer agent's
wealth. Fig. 1 shows that the distance in wealth between the richest and the poorest is huge. Practically, the entire wealth of  society concentrates
in the hand of very few people. In fact, even if both agents in a
transaction have the same chance to win or lose, the risk for the rich
trader is smaller because if he/she loses, the loss is a smaller fraction of
his/her own wealth  than that which the poorer
agent may lose. Consequently, there is a high probability in this model that
a very few people accumulate almost the entire wealth available and the
others become extremely poor, as Fig. 1 shows. An ideal society that adopts
such a trading policy will ultimately collapse.

\textbf{Random asymmetric trade price.} The third model assumes that the
poorer of the two traders is slightly advantaged in the trade, and has a
greater chance to formulate a good deal. In practice, we implement this
ability to make a good deal by assuming that if $w_{p}$ and $w_{r}$ are the
wealth of the poorer and the richer of the two traders, respectively, the
probability for the poorer trader to profit is 
\begin{equation}
\Pi =0.5+f~\frac{w_{r}-w_{p}}{w_{r}+w_{p}}~,  \label{prowin}
\end{equation}
where $0\leq f\leq 0.5$ is the asymmetry flux index. Eq. (\ref{prowin})
assumes that if the two agents have the same degree of wealth, they have the
same chance to get a good deal, while if one trader is much richer than the
other, the poorer is advantaged with the maximum probability of $\Pi =0.5+f$.
\begin{figure}
\epsfig{file=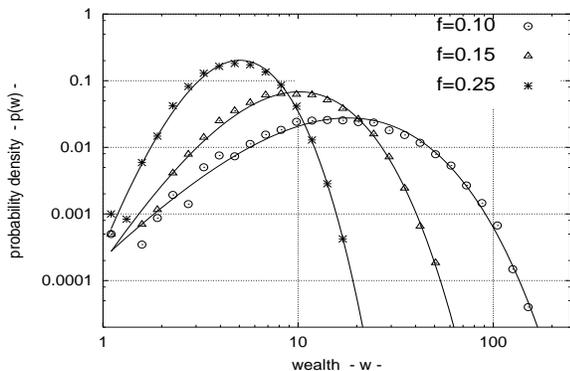,height=8cm,width=5cm,angle=-90}
\caption{Asymmetric, stationary economy. $f$ is the flux index. The wealth is in  units of the poorest agent's wealth.  }
\end{figure}
In the computer simulations shown in Fig. 2, we calculate three different
situations with $f=0.10$,  $f=0.15$ and  $f=0.25$, and the fraction
of wealth that may move in the transaction is assumed to vary between 0\% and 10\% of the
poorer agent's wealth. We determine that these computer-generated wealth pdf, as Fig. 2 shows,
are very well fit by curves of the form 
\begin{equation}
p(w)=a~w^{\gamma }\exp (-~b~w^{\delta })~.  \label{fitt1}
\end{equation}

Fig. 2 shows that with this model the economy of a society does not collapse
as it did in the previous case. The random asymmetric trade price model
yields a stable distribution of wealth, $p(w)$, well fitted by (\ref{fitt1}). 
The present model leads to a partitioning of society into three parts:
poor, middle and rich classes,  where the middle is the largest class
followed by the poor and finally the rich. The economic gap between richest
and poorest is not unrealistically wide, as it was in the random symmetric
model. If the curves of Fig. 2 seem unrealistic, it is because the gap in
the wealth between the richest and poorest is too small! However, the figure
shows that the separation in wealth between the rich and poor increases by
decreasing the asymmetric flux index $f$ and a more realistic distribution
with an exponential-like tail 
may emerge by decreasing this parameter.

In summary, the random asymmetric trade price model  produces a
stable wealth pdf that may be fitted by functions of the type of Eq. 
(\ref{fitt1}). The only difficulty with this model is that it does not reproduce
many experimental data, like those analyzed by Pareto, which suggest an
inverse power-law distribution that is not fitted by Eq. (\ref{fitt1}). We
shall show that this failure is due to the assumption that society's total
wealth remains constant. 

\textbf{Random asymmetric trade price with nonstationary wealth.} Our last
computational model  implements the mechanism of the
previous model with the additional  property that the total wealth is not
constant. The wealth of each individual trader is subject to a random
destruction-creation dynamics, due to the natural ebb and flow of  wealth
through human creativity,  work and investment. 
We introduce the non-stationary mechanism through a risk index $r>0$, that
measures the standard deviation of the destruction-creation wealth process.
In our computer simulation we assume that after every 10,000 transactions
the wealth of each trader is reinitialized by the following expression 
\begin{equation}
w_{i}(t+1)=(1+r~\xi )~w_{i}(t)~,  \label{rinizi}
\end{equation}
where $w_{i}(t)$ is the wealth of the i-$th$ agent after the t-$th$ epoch
and $w_{i}(t+1)$ is that agent's wealth at the start of the new epoch, $t+1$.
Finally,  $\xi$ is a zero-centered  Gaussian random
variable with unit variance. The rare instances of  too negative a random value of the variable $\xi$, that makes the wealth $w(i)$ negative, are neglected.

Fig. 3a shows the distribution of wealth for three computer simulations
obtained by assuming a fixed risk index $r=0.1$ and varying the asymmetry
flux index $f$:  $f=0.10$, $f=0.15$ and $f=0.30$. Fig. 3b shows another three
computer simulations obtained by assuming a fixed asymmetry flux index $f=0.1$ 
and varying the risk index, $r=0.05$, $r=0.10$ and $r=0.15$. The fraction
of wealth that may move between agents in a single  transaction is assumed again to vary between 0\% and 10\% of the poorer agent's wealth. Finally, Figs. 3a and 3b show that the computer-generated wealth pdf
can be well fitted for a very large region by curves of the form 
\begin{equation}
p(w)=a~w^{\gamma }/(1+b~w)^{\gamma +\delta }~.  \label{fitt2}
\end{equation}

As in the previous model, we obtain a stable wealth distribution $p(w)$ that partitions society into the three classes. The
wealth gap between the richest and poorest increases by decreasing the
asymmetric flux index $f$ and/or by increasing the risk index $r$. 

We stress that Eq. (\ref{fitt2}) is characterized by an inverse power-law tail of the
type $p(w)\propto 1/w^{\delta }$, where $\delta $ is the inverse power-law
exponent. This inverse power-law distribution is compatible with Pareto's
measurements. In fact, Fig. 4 shows the cumulative probability $P(w)$ 
for an economy with $f=0.15$ and $r=0.075$, fitted with the Pareto
distribution and having a Pareto exponent of $\alpha =\delta -1=1.48$.
This figure is consistent with the fit to U.S.  income data
made in Figure 35 of Ref.\cite{montroll}.
\begin{figure}
\epsfig{file=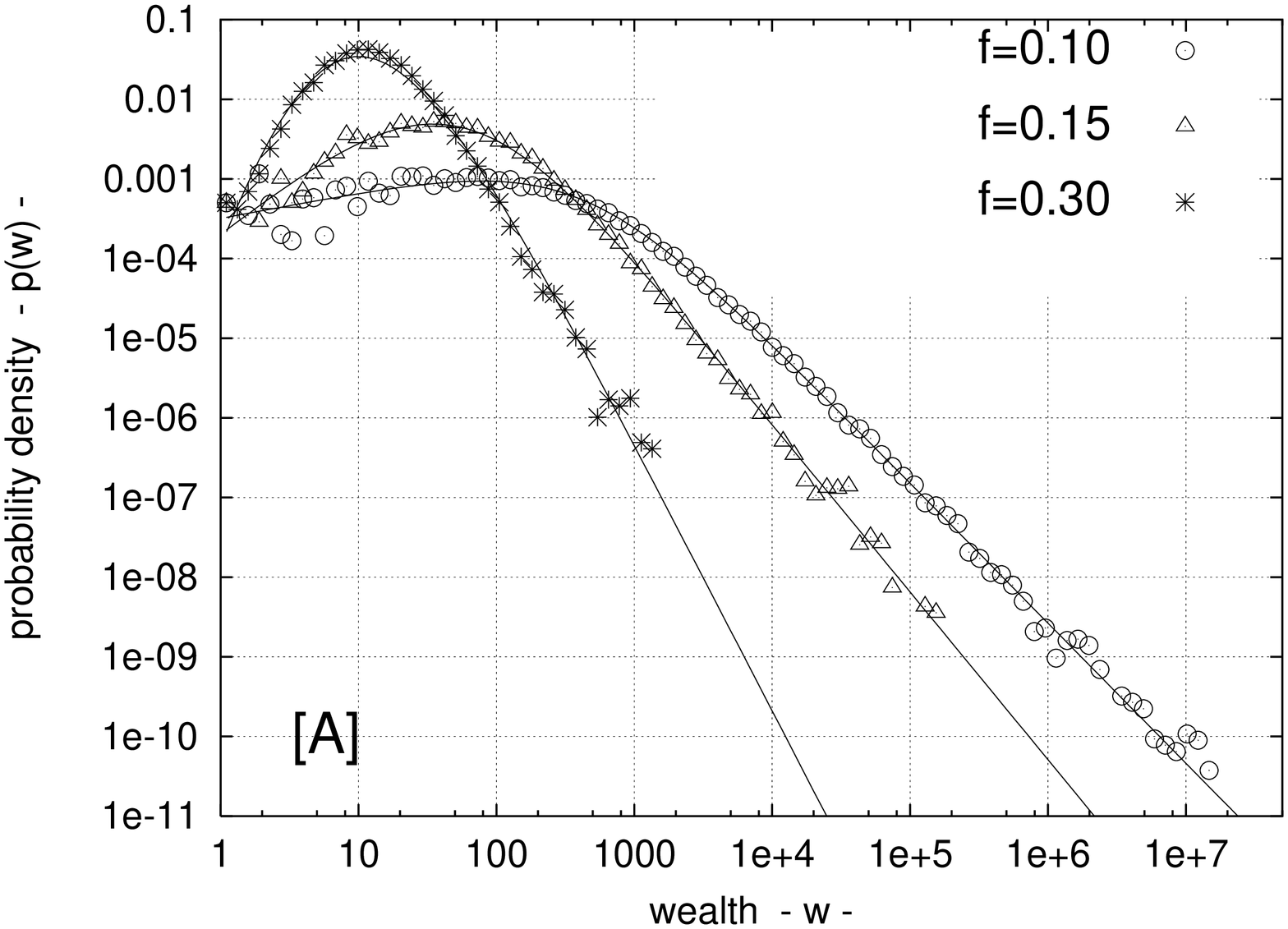,height=8cm,width=5cm,angle=-90}

\epsfig{file=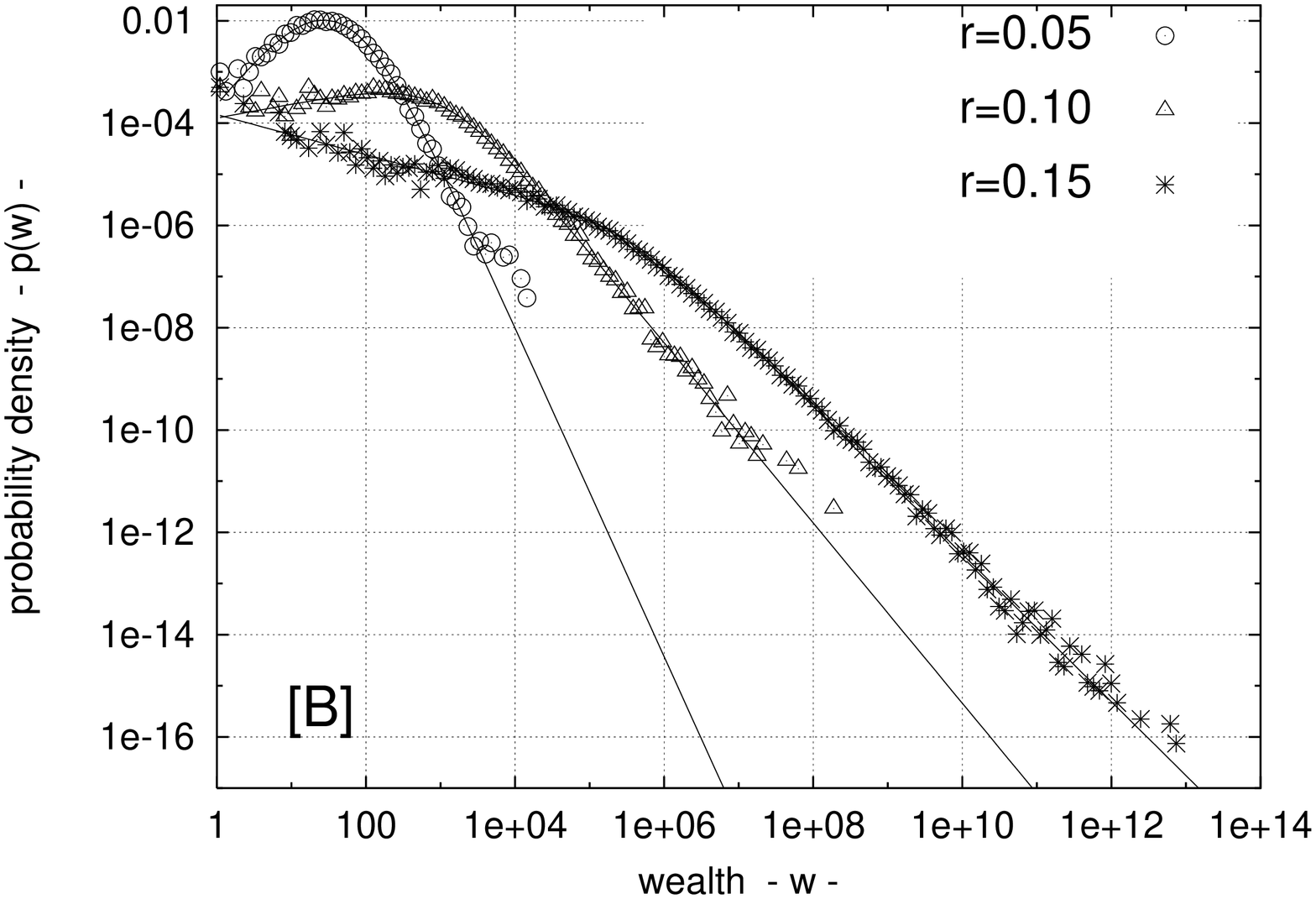,height=8cm,width=5cm,angle=-90}
\caption{Asymmetric, non-stationary economy. (a) The risk index is $r=0.1$; the fit gives: $\delta=1.76\pm0.05$,  $\delta=2.1\pm0.05$, $\delta=3.3\pm0.2$. (b)The flux index is $f=0.1$; the fit gives: $\delta=3.2\pm0.1$,  $\delta=1.76\pm0.03$, $\delta=1.44\pm0.02$. The wealth is measured in  units of the poorest agent's wealth.  }
\end{figure}

A society that allows a small class of people to absorb all its wealth is
unstable and will either collapse economically, or be destroyed by revolution. This may
also happen in the absence of economic exploitation of the poor by the rich,
as we explained above. A stable society requires that the poor have an
advantage in transactions with the wealthy and are protected by particular
rights and marketing freedom. On the other hand, in a real human society,
wealth does not only move randomly from one individual to another through
trades, but is continuously created and destroyed through an individual's
work, creativity and investment. The human aspiration of personal riches
requires a positive risk index $r$ that generates both a large middle class
and a small rich class, when applied across the society. The asymmetry flux
index $f$ and the risk index $r$ express  the human need to share and  the human need to create, respectively. Therefore, we conclude from the model that both mechanisms must be present to some degree in any human society that functions effectively and remains stable. In fact, when these two mechanisms are present, according to our model, the wealth pdf assumes the inverse power-law
distribution of Pareto at the high income end, and, more realistically, still retains a small but
finite population at the low income end. 

\begin{figure}
\epsfig{file=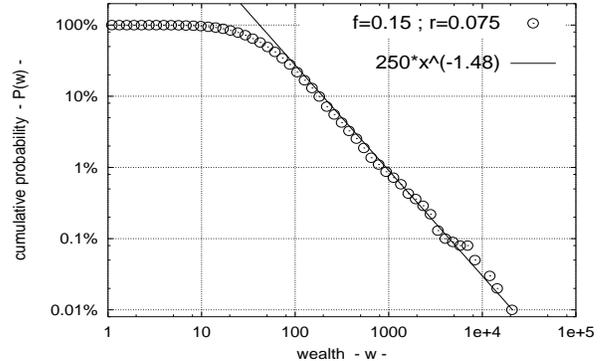,height=8cm,width=5cm,angle=-90}
\caption{Cumulative probability for an asymmetric, non-stationary economy. $f=0.15$, $r=0.075$. The Pareto's exponent is $\alpha=1.48\pm0.02$. }
\end{figure}

{ \textbf{Acknowledgment: }}
N.S. thanks the Army Research Office for support under grant DAAG5598D0002.

\end{document}